\documentclass[aps,pre,reprint,superscriptaddress,amsmath,amssymb]{revtex4-2}
\usepackage{graphicx}
\usepackage{tabularx}
\usepackage{dcolumn}
\usepackage[hypertexnames=false]{hyperref}
\usepackage{txfonts}
\usepackage{color}
\usepackage{bm}
\usepackage{here}
\usepackage{siunitx}

\newcommand{\sectionprl}[1]{{\em #1}\/.---}

\usepackage{booktabs}
\newcolumntype{C}[1]{>{\centering\arraybackslash}p{#1}}

\newcommand{\noneq}{\rm neq}
\newcommand{\eq}{\rm eq}

\begin{document}
\title{Long-Range Correlations under Temperature Gradients:\\ A Molecular Dynamics Study of Simple Fluids}
\author{Hiroyoshi Nakano}
\affiliation{Institute for Solid State Physics, University of Tokyo, Kashiwa, Chiba 277-8581, Japan}
\author{Kazuma Yokota}
\affiliation{Department of Physics, Kyoto University, Kyoto 606-8502, Japan}

\date{\today}
\begin{abstract}
In fluids under temperature gradients, long-range correlations (LRCs) emerge generically, leading to enhanced density fluctuations.
This phenomenon, characterized by the $\bm{q}^{-4}$ divergence in the static structure factor (where $\bm{q}$ is the wavenumber), has been extensively studied both theoretically and experimentally.
However, they remain unexplored in Hamiltonian particle systems using molecular dynamics (MD) simulations.
This Letter reports the first MD study to provide unambiguous observations of the LRCs.
We demonstrate this by three distinct approaches: (1) measuring the static structure factor and directly observing the $\bm{q}^{-4}$ divergence characterizing the LRCs; (2) detecting the corresponding $\bm{q}^{-4}$ divergence in the dynamic structure factor; (3) establishing a quantitative agreement between MD results and predictions from fluctuating hydrodynamics, the phenomenological theory that predicts the LRCs.
Our findings demonstrate that MD simulations offer a powerful complementary tool to theoretical and experimental investigations of LRCs.
\end{abstract}
\maketitle

\sectionprl{Introduction}
In equilibrium systems, the spatial correlations of fluctuations typically decay exponentially with a finite correlation length. 
Long-range correlations (LRCs), characterized by algebraic decay, are restricted to exceptional cases, such as critical points of second-order phase transitions and ordered states with continuous symmetry breaking.
Such singular behaviors of fluctuations have been extensively studied as a central theme in equilibrium statistical mechanics.
In contrast, divergent correlation lengths and LRCs are ubiquitous in nonequilibrium steady states~\cite{Garrido1990-dy, Schmittmann1995-mc, Dorfman1994-cl, De_Zarate2006-xw}.
Indeed, nonequilibrium LRCs are considered to appear in a variety of nonequilibrium systems.
The typical examples include lattice gases subjected to driving forces~\cite{Garrido1990-dy, Derrida1993-ep, Schmittmann1995-mc, Schmittmann1998-cg, Praestgaard2000-bj}, fluids under temperature gradients~\cite{Dorfman1994-cl, De_Zarate2006-xw, Vailati2012-yz, Bedeaux2015-lu, Croccolo2016-mn, Kirkpatrick1982-dq, Ronis1982-uk, Law1988-zb, Law1990-pg, Segre1992-cd, Li1994-rh, Li1994-hv, Li1994-vw, Vailati1996-dw, Ortiz_de_Zarate2001-ho, Ortiz_de_Zarate2001-ab, Takacs2008-ax, Takacs2011-gg, Vailati2012-yz, Kirkpatrick2013-ic, Kirkpatrick2016-xt, Kirkpatrick2021-rv, Kirkpatrick2024-ry}, fluids under shear flow~\cite{Onuki1979-ji, Lutsko1985-zb, Wada2003-je, Otsuki2009-ld, Ortiz-de-Zarate2011-ec, Varghese2017-vw, Ortiz_de_Zarate2019-pp, Nakano2022-kv}, liquid mixtures under concentration gradients~\cite{Dorfman1994-cl, De_Zarate2006-xw, Vailati2012-yz, Bedeaux2015-lu, Croccolo2016-mn, Vailati1997-ez, Brogioli2001-ct, Ortiz-de-Zarate2004-vb, Vailati2006-zc, Mazzoni2006-ji, Croccolo2006-bm, Vailati2011-rz, Cerbino2015-jh, Croccolo2016-et, Kirkpatrick2015-rg, Ortiz_de_Zarate2015-pd, Kirkpatrick2016-ev, Donev2011-hf, Donev2011-vk, BalboaUsabiaga2012-sh, Balakrishnan2014-vg, Srivastava2023-nx, Castellini2023-yo}, and active matter systems with anisotropy~\cite{Adachi2022-pt, Nakano2024-th, Adachi2024-cx}.
The existence of nonequilibrium LRCs serves as a prime example of the rich behaviors of nonequilibrium systems beyond the equilibrium statistical mechanics, which continues to attract considerable attention~\cite{Peraud2017-xt, Mahdisoltani2021-ja, Doyon2023-wn, Hubner2024-xc}.

The exploration of nonequilibrium LRCs has mainly been carried out using kinetic theory~\cite{Kirkpatrick1982-dq}, fluctuating hydrodynamics~\cite{Schmittmann1995-mc, De_Zarate2006-xw}, and driven lattice gas models~\cite{Garrido1990-dy, Derrida1993-ep}.
However, beyond these mesoscopic approaches and toy models, only a handful of instances have been directly demonstrated in real experiments or microscopic Hamiltonian particle systems.
Improving this situation is crucial not only for advancing our understanding of nonequilibrium LRCs but also for discussing their implications for the nonequilibrium properties of real materials.

The aim of this Letter is to directly observe LRCs appearing in nonequilibrium fluids under temperature gradients~\cite{Kirkpatrick1982-dq, Ronis1982-uk, Law1988-zb, Law1990-pg, Segre1992-cd, Li1994-rh, Li1994-hv, Li1994-vw, Vailati1996-dw, Ortiz_de_Zarate2001-ho, Ortiz_de_Zarate2001-ab, Takacs2008-ax, Takacs2011-gg, Vailati2012-yz, Kirkpatrick2013-ic, Kirkpatrick2016-xt, Kirkpatrick2021-rv, Kirkpatrick2024-ry} by performing large-scale molecular dynamics (MD) simulations of a Hamiltonian particle system.
Fluids subjected to external gradients, such as temperature or concentration gradients, serve as prominent examples where nonequilibrium LRCs have been successfully observed experimentally~\cite{Bedeaux2015-lu}.
While nonequilibrium LRCs have also been reported in isothermal systems under concentration gradients~\cite{Vailati1997-ez, Brogioli2001-ct, Castellini2023-yo}, this work specifically focuses on temperature gradients.
LRCs under both types of gradients have been extensively studied theoretically and experimentally and are understood to originate from a common underlying mechanism —the coupling between the gradient and velocity fluctuations parallel to the gradient~\cite{De_Zarate2006-xw}—.
For temperature gradients, LRCs were first theoretically predicted in the early 1980s~\cite{Kirkpatrick1982-dq, Ronis1982-uk} and subsequently confirmed through experimental observations around 1988~\cite{Law1988-zb, Law1990-pg, Segre1992-cd}.
In these experiments, a nonequilibrium correction term of the time-dependent structure factor was observed and evidence of the LRCs was found from its wave number dependence.
Recent experimental advances~\cite{Takacs2008-ax, Takacs2011-gg, Vailati2006-zc, Mazzoni2006-ji, Croccolo2006-bm, Vailati2011-rz, Cerbino2015-jh, Croccolo2016-et} have further confirmed these findings by eliminating the influence of gravity under microgravity conditions and achieving unambiguous observations of the static structure factor~\cite{Takacs2011-gg}.
The recent experimental studies are summarized in Ref.~\cite{Vailati2012-yz, Bedeaux2015-lu, Croccolo2016-mn}.

We here present MD simulation results for dense liquids, focusing on two-dimensional (2D) systems.
The 2D simulations allow us to access significantly larger system sizes than are typically feasible in three dimensions.
Critically, when analyzed in wave-vector space as in this study, the fundamental physics of LRCs remains unchanged qualitatively between two and three dimensions.
The properties of LRCs induced by concentration gradients in 2D systems have been discussed in detail in Ref.~\cite{Brogioli2017-hf}.
In addition, our MD simulations do not include the effects of gravity, which are known to suppress LRCs in long wavelength regimes~\cite{Segre1993-fp, Takacs2008-ax}.
These effects are expected to be negligible for length scales accessible in MD simulations.
Also, recent experimental advances in microgravity environments have enabled direct observation of LRCs without the influence of gravity.

We demonstrate that the large-scale MD simulations, with $1$ to $10$ million particles, successfully reproduce all of the evidence for the LRCs previously observed in experiments.
Although strong finite-size effects, consistent with previous studies~\cite{Ortiz_de_Zarate2001-ho, Ortiz_de_Zarate2001-ab, Mazzoni2006-ji, Croccolo2006-bm, Takacs2011-gg, Vailati2011-rz, Ortiz_de_Zarate2015-pd, Cerbino2015-jh}, hinder the observation of the LRCs and require the large-scale simulations, such simulations are achievable with current computational resources.
Thus, our results suggest that MD simulations can be a powerful tool for exploring LRCs in realistic nonequilibrium situations.
While this Letter primarily focuses on dense liquids, we also performed the MD simulations of dilute gases.
See the Supplemental Material (SM)~%
\footnote{
See Supplemental Material for additional information on (1) the dilute gas simulation, (2) the simulation results under large temperature gradients, (3) the simulation details and observation protocols, and (4) the measurement of fluid properties.
}
for the results, which show that identical results hold for dilute gases as well, revealing the generality of our findings.

\begin{figure}[tb]
\begin{center}
\includegraphics[scale=1]{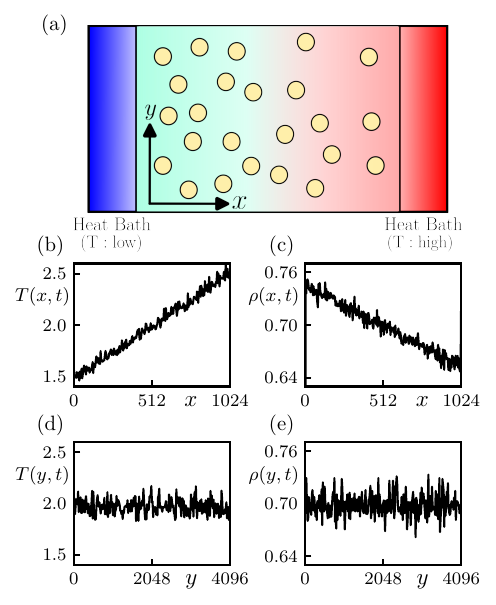}
\caption{
(a) Schematic illustration of the system.
The $x$ axis is taken along the direction of the temperature gradient.
(b)-(e) Typical one-dimensional temperature and density profiles in the nonequilibrium steady state, observed in the MD simulation.
The setup is the same as that of the red curve in Fig.~\ref{fig2}~(b), which will be explained later.
}
\vspace{-0.5cm}
\label{fig1}
\end{center}
\end{figure}
\sectionprl{Basic properties of long-range correlations}
Before proceeding to the main part of this Letter, we briefly review the LRCs appearing in fluids under temperature gradients.
Let us consider a fluid with a uniform temperature gradient along the $x$ axis, as illustrated in Fig.~\ref{fig1}~(a).
To aid in understanding the phenomenon of LRCs, we first present one-dimensional temperature and density profiles in Figs~\ref{fig1}~(b)-(e).
These profiles, observed in the MD simulation (described later), illustrate fluctuations of temperature and density around their steady-state profiles at a given time $t$.
Here, one-dimensional profiles are obtained from the two-dimensional temperature and density fields, $T(\bm{r}, t)$ and $\rho(\bm{r}, t)$, by integrating along the $x$ or $y$ direction, for example, $T(x, t) := \int dy T(\bm{r}, t) / L_y$ or $T(y, t) := \int dx T(\bm{r}, t) / L_x$.
The temperature difference at the boundaries induces a nearly uniform temperature gradient in $T(x, t)$, which in turn leads to a density gradient in $\rho(x, t)$.
Moreover, we can clearly observe fluctuations around these steady-state profiles.

The dynamics of such fluctuations at the mesoscopic scale is considered to be effectively captured by fluctuating hydrodynamics, a continuum theory that incorporates thermal fluctuations into the hydrodynamic equations~\cite{Landau1959-eu, BalboaUsabiaga2012-sh, Balakrishnan2014-vg, Srivastava2023-nx}.
The FH framework has been the subject of efforts to derive it from microscopic particle systems, for instance, using projection operator methods.
With appropriately chosen parameters, it is expected to provide a quantitative prediction of dynamical phenomena.
In particular, it has been used to predict fluctuation behavior under temperature gradients~\cite{De_Zarate2006-xw}.
Here, based on this framework, we explain the basic nature of the LRCs in the presence of a temperature gradient.
A detailed review of fluctuating hydrodynamics, including the key results used in this Letter, is summarized in Ref.~\cite{De_Zarate2006-xw}.

The main focus of this Letter is density fluctuations.
We then calculate the density-density correlation function $\langle \rho(\bm{r},t) \rho(\bm{r}',t) \rangle^{\rm th}_{\rm ss}$ or equivalently, the static structure factor $S^{\rm th}(\bm{q})$, defined by
\begin{align}
    \langle \rho(\bm{q},t) \rho(\bm{q}',t) \rangle^{\rm th}_{\rm ss} = \rho_0 S^{\rm th}(\bm{q}) (2\pi)^2\delta(\bm{q}+\bm{q}'),
\end{align}
where $\rho(\bm{r},t)$ is the density field, $\rho(\bm{q},t)$ its Fourier transform, and $\rho_0$ the averaged density.
Here, we use the superscript "th" to denote quantities obtained from theoretical discussions.

In equilibrium, $\langle \rho(\bm{r},t) \rho(\bm{r}',t) \rangle^{\rm th}_{\rm ss}$ decays exponentially with a correlation length on the microscopic scale.
On the hydrodynamic scale, this decay is described by a delta function:
\begin{align}
    \langle \rho(\bm{r},t) \rho(\bm{r}',t) \rangle^{\rm th}_{\rm ss} = \rho_0 S^{\rm th}_{\eq}\delta(\bm{r}-\bm{r}') \ \ \ {\rm with}  \ \ S^{\rm th}_{\eq}:= \frac{k_B \alpha_p^2 T}{c_p-c_v}.
\end{align}
Here, $T$ is the temperature, $c_p$ the specific heat capacity at constant pressure, $c_v$ the specific heat capacity at constant volume, and $\alpha_p$ the thermal expansion coefficient.
The corresponding static structure factor $S^{\rm th}(\bm{q})$ is then given by:
\begin{align}
    S^{\rm th}(\bm{q}) = S^{\rm th}_{\eq}.
\end{align}

\begin{figure*}[tb]
\begin{center}
\includegraphics[scale=1.0]{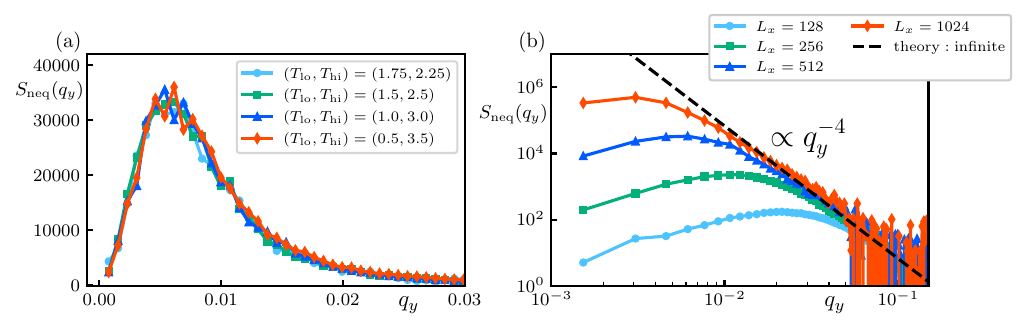}
\end{center}
\vspace{-0.5cm}
\caption{
The MD results of the nonequilibrium correction of the static structure factor $S_{\noneq}(q_y) := (S(q_y)-S_{\mathrm{eq}}) / (\nabla T)_0^2$. 
(a) Dependence of $S_{\noneq}(q_y)$ on the temperature gradients $(\nabla T)_0$.
We fix the system size at $(L_x, L_y)=(512, 8192)$ and vary the temperatures at the two ends as $(T_{\mathrm{left}}, T_{\mathrm{right}})=(1.75, 2.25)$, $(1.5, 2.5)$, $(1.0, 3.0)$, $(0.5, 3.5)$, where the midpoint temperature is fixed at $T_{\mathrm{mid}}:=(T_{\mathrm{left}}+T_{\mathrm{right}})/2=2.0$.
The MD results are not influenced by the value of $(\nabla T)_0$.
(b) Dependence of $S_{\noneq}(q_y)$ on the system size $L_x$.
Data is presented on a log-log scale.
We maintain the temperatures at the two ends at $(T_{\mathrm{left}}, T_{\mathrm{right}})=(1.5, 2.5)$ and vary the system size as $(L_x, L_y)=(128, 8192)$, $(256, 8192)$, $(512, 8192)$, $(1024, 8192)$.
The blue curve presents the same data as the green curve in panel (a), but plotted on a log-log scale.
The black dashed line represents the theoretical prediction [Eq.~(\ref{eq: noneq part of stfactor})] for an infinitely large system.
The MD results approach the theoretical prediction as the wall distance $L_x$ increases.
}
\label{fig2}
\end{figure*}
Fluctuating hydrodynamics predicts that $S^{\rm th}(\bm{q})$ is enhanced due to the temperature gradient $(\nabla T)_0$, which is expressed to leading order as follows:
\begin{align}
S^{\rm th}(\bm{q}) = S^{\rm th}_{\mathrm{eq}} + (\nabla T)_0^2 S^{\rm th}_{\mathrm{neq}}(\bm{q}),
\label{eq: stfactor infinite / theory}
\end{align}
where the nonequilibrium correction is proportional to $(\nabla T)_0^2$.
For an infinitely large system, where boundary effects can be neglected, the explicit expression for $S^{\rm th}_{\mathrm{neq}}(\bm{q})$ is given by:
\begin{align}
    S^{\rm th}_{\mathrm{neq}}(\bm{q}) &= A_{\mathrm{neq}}\frac{\hat{q}_y^2}{\bm{q}^4} \quad \textrm{with} \quad A_{\mathrm{neq}} = \frac{\alpha_p^2 k_B T_{\mathrm{mid}}}{a_T (\nu + a_T)},
    \label{eq: noneq part of stfactor}
\end{align}
where $T_{\mathrm{mid}}$ is the temperature at the center of the system, $\hat{q}_y:=q_y/|\bm{q}|$, $\nu$ the kinetic viscosity, and $a_T$ the thermal diffusivity.
This $\bm{q}^{-4}$ divergence in $S^{\rm th}_{\noneq}(\bm{q})$ is a direct manifestation of the LRCs.
Interestingly, it is much stronger than the $\bm{q}^{-2}$ divergence of critical density fluctuations at a liquid-vapor critical point.

The corresponding real-space correlation function is:
\begin{align}
    \frac{\langle \rho(t, \bm{r}) \rho(t, \bm{r}')\rangle^{\rm th}_{\rm ss}}{\rho_0}= S^{\rm th}_{\eq}\delta(\bm{r}-\bm{r}') + (\nabla T)_0^2 C^{\rm th}_{\mathrm{neq}}(\bm{r},\bm{r}').
\end{align}
The nonequilibrium contribution  $C^{\rm th}_{\mathrm{neq}}(\bm{r}, \bm{r}')$ has a complicated form; for example, for $x=x'$ and $y,y'\ll L_y$, it is:
\begin{align}
    &C^{\rm th}_{\mathrm{neq}}(\bm{r}, \bm{r}') \nonumber \\ &= \frac{A_{\mathrm{neq}}L_y^2}{32\pi^3} \left[ 1 + 3 \Big|\frac{y-y'}{L_y/\pi}\Big|^2 \log \Big|\frac{y-y'}{L_y/\pi} \Big| - B \Big|\frac{y-y'}{L_y/\pi}\Big|^2 + \cdots \right]
\end{align}
with $B:=(11-12\gamma)/4$ ($\gamma$: Euler's constant).
Notably, this real-space correlation function lacks a characteristic decay length and its magnitude is proportional to $L_y^2$.
This manifests that the temperature gradient significantly enhances density fluctuations, resulting in the LRCs that persist across the entire system.

\sectionprl{Setting of MD simulations}
We consider a system of $N$ particles confined within a rectangular box of dimensions $[-0.5L_x, 0.5L_x] \times [-0.5L_y, 0.5L_y]$ (see Fig.~\ref{fig1} (a)).
These particles interact via the Weeks--Chandler--Andersen (WCA) potential, given by 
\begin{align}
V_{\mathrm{WCA}}(r) =
\begin{cases}
4\epsilon \Big[ \left( \frac{\sigma}{r} \right)^{12} - \left( \frac{\sigma}{r} \right)^6 + \frac{1}{4} \Big], & r \leq 2^{1/6}\sigma \\
0, & {\mathrm{otherwise}},
\end{cases}
\end{align}
where $\epsilon$ and $\sigma$ represent the well depth and atomic diameter, respectively.
All particles have the same mass $m$.
The particles are confined by walls at $x = -0.5L_x$ and $x = 0.5L_x$, while the periodic boundary condition is applied along the $y$ axis.
To establish a temperature gradient along the $x$ axis, we apply Langevin thermostats in the regions $[-0.5L_x, -0.48L_x]$ and $[0.48L_x, 0.5L_x]$.
The temperatures of these thermostats are, respectively, set to $T_{\mathrm{left}}$ and $T_{\mathrm{right}}$ ($T_{\mathrm{left}} < T_{\mathrm{right}}$), as illustrated in Fig.~\ref{fig1} (a).
Our analysis focuses on the bulk region, $\mathcal{B} := [-0.48L_x, 0.48L_x] \times [-0.5L_y, 0.5L_y]$, which excludes the thermostatted zones.
Note that in the hydrodynamic description, the thermostat regions are considered as the effective walls, and the wall positions are located at $x = \pm 0.48L_x$, not $x = \pm 0.5L_x$ (See SM~\footnotemark[1] for this validity).

We measure physical quantities in units of energy $\epsilon$, length $\sigma$, and time $\tau = \sigma \sqrt{m/\epsilon}$.
All the MD simulations are performed by LAMMPS~\cite{Thompson2022-eu}.
We fix the density to $\rho=0.7$ and the temperature at the center of the system to $T_{\mathrm{mid}}:=(T_{\mathrm{left}}+T_{\mathrm{right}})/2=2.0$.
See SM~\footnotemark[1] for observation protocols and the thermodynamic and transport properties of the fluid.

\begin{figure*}[tb]
\begin{center}
\includegraphics[scale=1.0]{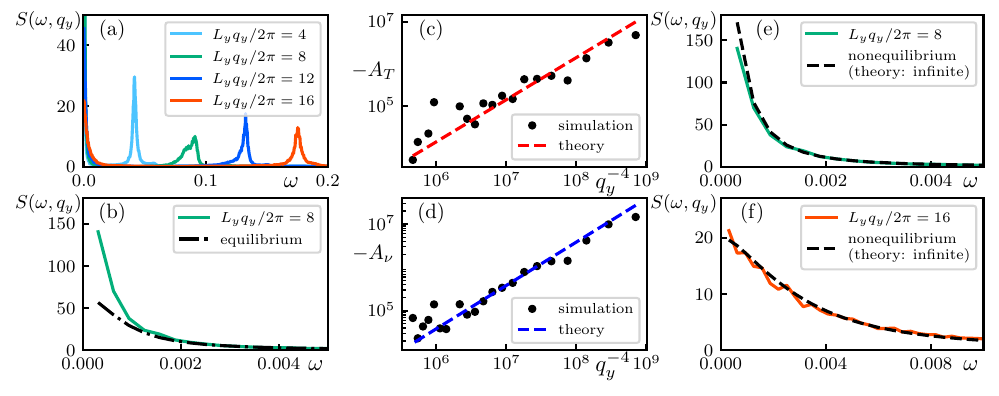}
\end{center}
\vspace{-0.6cm}
\caption{
The MD results of the dynamic structure factor $S(\omega, q_y)$.
The parameters are chosen as $(L_x, L_y)=(1024, 4096)$ and $(T_{\mathrm{left}}, T_{\mathrm{right}})=(1.5, 2.5)$, corresponding to the red line in Fig.~\ref{fig1}~(b).
(a) Rayleigh and Brillouin peaks of $S(\omega, q_y)$ for several wavenumbers, $q_y$. 
(b) $S(\omega, q_y)$ near the Rayleigh peak for $q_y/2\pi=8/L_y$.
The colored solid curve represents an enlarged view of the data in panel (a), while the black dot-dashed curve represents the corresponding equilibrium result.
Nonequilibrium effects dominate the behavior at small $\omega$.
(c, d) The $q_y^{-4}$ dependence of $A_T(q_y)$ and $A_{\nu}(q_y)$.
The colored dashed lines represent the theoretical predictions [Eq.~(\ref{eq: expression for Anu and AT})] for an infinitely large system.
The MD results agree accurately with the theoretical predictions, revealing that $A_T(q_y)$ and $A_{\nu}(q_y)$ exhibit the $q_y^{-4}$ divergence.
(e, f) Comparison of the Rayleigh peak of $S(\omega, q_y)$ with theoretical predictions for $q_y/2\pi = 8/L_y$ and $q_y/2\pi = 16/L_y$.
The colored solid lines represent the MD results (in particular, the data in panel (e) are replotted from panel (b)).
The black dashed curves represent the theoretical predictions under the temperature gradient for an infinitely large system.
We observe excellent agreement between the MD results and theoretical predictions.
}
\label{fig3}
\end{figure*}
\sectionprl{Direct observation of long-range correlations}
The LRCs can be directly observed through the static structure factor $S(\bm{q})$.
In our setup, it is defined as:
\begin{align}
    \langle \rho(t, \bm{q}) \rho(t, \bm{q}')\rangle_{\rm ss} = \rho_0 S(\bm{q}) (2\pi)^{2} \delta(\bm{q} + \bm{q}'),
    \label{eq: def of stfactor}
\end{align}
where $\rho(t, \bm{q}) := \int_{\mathcal{B}} d^2 \bm{r} \rho(t, \bm{r}) e^{-i\bm{q} \cdot \bm{r}}$, $\rho_0$ is the average density in the bulk region $\mathcal{B}$, and $\bm{q}=(2\pi n/(0.96L_x), 2\pi m/L_y)$ for $n, m \in \mathbb{Z}$.
In this study, we focus on the case of $q_x = 0$, specifically, $S(q_y):= S(q_x=0, q_y)$, which is readily accessible through small-angle scattering experiments.

To analyze the nonequilibrium correction to the static structure factor, we decompose $S(q_y)$ into an equilibrium contribution and a nonequilibrium correction, as suggested by Eq.~(\ref{eq: stfactor infinite / theory}):
\begin{align}
S(q_y) = S_{\eq} + (\nabla T)_0^2 S_{\noneq}(q_y).
\label{eq: formal expansion of stfactor}
\end{align}
Figure~\ref{fig2} presents the MD results for $S_{\noneq}(q_y)$.
The typical instantaneous temperature and density profiles are shown in Fig.~\ref{fig1}~(b)-(e).

We first investigate the dependence of $S_{\noneq}(q_y)$ on the applied temperature gradient $(\nabla T)_0$.
The results of this analysis are shown in Fig.~\ref{fig2}~(a). 
We observe that $S_{\noneq}(q_y)$ collapses onto a single universal curve over a wide range of temperature gradients, indicating that $S_{\noneq}(q_y)$ is independent of $(\nabla T)_0$.
This finding is noteworthy because the temperature dependence of the fluid properties, such as the thermal conductivity, could potentially influence $S_{\noneq}(q_y)$, as predicted by fluctuating hydrodynamics in Eq.~(\ref{eq: noneq part of stfactor}).
Indeed, using argon parameters for the WCA potential, our system [red curve in Fig.~\ref{fig2}~(a)] experiences a temperature gradient as large as $360\textrm{K}$ over $167\textrm{nm}$.
We note that such a temperature gradient (360 K over 167 nm) is unrealistically large for typical experiments with argon.
However, even under such extreme, hypothetical conditions, $S_{\noneq}(q_y)$ shows negligible dependence on $(\nabla T)_0$ and is effectively characterized by the midpoint temperature $T_{\rm mid} = 2.0$.
We find that this robustness persists even when larger temperature gradients are applied and both thermodynamic and transport coefficients exhibit significant temperature dependence.
We discuss this point in SM~\footnotemark[1] for details.

We now turn our attention to the $q_y$ dependence of $S_{\noneq}(q_y)$.
Figure~\ref{fig2}~(b) shows the behavior of $S_{\noneq}(q_y)$ as we systematically vary the wall separation $L_x$ while keeping the temperatures at the ends, $(T_{\mathrm{left}}, T_{\mathrm{right}})$, constant.
The blue curve presents the same data as the green curve in Fig.~\ref{fig2}~(a), but plotted on a log-log scale.
The MD results are compared against the theoretical prediction for $S^{\rm th}_{\noneq}(q_y)$ [Eq.~(\ref{eq: noneq part of stfactor})].
Note that this prediction is for an infinitely large system.
For comparison, no fitting parameters were used; the parameters in Eq.~(\ref{eq: noneq part of stfactor}) were determined from the separate MD simulation~\footnotemark[1].
As $L_x$ increases, the MD results converge towards the theoretical prediction, clearly revealing the emergence of the $q_y^{-4}$ divergence [Eq.~(\ref{eq: noneq part of stfactor})].
This direct observation of the $q_y^{-4}$ divergence confirms the successful detection of the LRCs in our MD simulation.

\sectionprl{Detection of long-range correlations from dynamical structure factor}
Historically, experimental detections of the LRCs under temperature gradients have relied on observations of the dynamic structure factor $S(\omega, \bm{q})$~\cite{Law1988-zb, Law1990-pg, Segre1992-cd, Li1994-rh, Li1994-hv, Li1994-vw}.
This quantity is defined as:
\begin{align}
    S(\omega, \bm{q}) = \int_{-\infty}^{\infty} \frac{dt}{2\pi} S(t, \bm{q})e^{-i\omega t} 
    \label{eq: def of dyfactor}
\end{align}
with $\langle \rho(t, \bm{q}) \rho(0, \bm{q}')\rangle_{\rm ss} = \rho_0 S(t, \bm{q}) (2\pi)^{2} \delta(\bm{q} + \bm{q}')$.
The static structure factor $S(\bm{q})$ corresponds to $S(t, \bm{q})$ at $t = 0$.

Figure~\ref{fig3}~(a) presents the MD results, showing the overall behaviors of $S(\omega, q_y)$ for various wavenumbers $q_y$.
For each $q_y$, $S(\omega, q_y)$ exhibits two types of peaks: Rayleigh and Brillouin peaks.
The Rayleigh peak is located near $\omega = 0$, while the Brillouin peaks are located near $\omega = \pm c_s q_y$, where $c_s$ is the speed of sound.
According to fluctuating hydrodynamics, temperature gradients primarily affect the Rayleigh peak.
Based on this prediction, we provide an enlarged view of $S(\omega, \bm{q})$ near the Rayleigh peak in Fig.~\ref{fig3}~(b), comparing the MD result under the temperature gradient with that for the equilibrium system.
As seen in this figure, $S(\omega, \bm{q})$ is enhanced by the temperature gradient for small $\omega$.
To identify the presence of the LRCs from this data, we implement two distinct protocols.

The first protocol is to detect, within $S(\omega, \bm{q})$, the counterpart to the $q_y^{-4}$ divergence observed in $S(\bm{q})$.
To understand the basis of this approach, we present the explicit expressions of $S(t, \bm{q})$ and $S(\omega, \bm{q})$ calculated from fluctuating hydrodynamics for an infinitely large system~\cite{De_Zarate2006-xw}:
\begin{align}
    S^{\rm th}(t, q_y) &= S^{\rm th}_{\mathrm{Ray}, \mathrm{eq}} \Big\{\left(1 + (\nabla T)_0^2 A_{T}(q_y)\right)e^{-a_T q_y^2 |t|} \nonumber \\
    &\hspace{3cm} - (\nabla T)_0^2 A_{\nu}(q_y)e^{-\nu q_y^2 |t|} \Big\}
    \label{eq: itfactor infinite} \\
    S^{\rm th}(\omega, q_y) &= S^{\rm th}_{\mathrm{Ray}, \mathrm{eq}}\Biggl\{\left(1 + (\nabla T)_0^2 A_T(q_y) \right)\frac{2a_T q_y^2}{\omega^2 + a_T^2 q_y^4} \nonumber \\
    &\hspace{2.5cm} - (\nabla T)_0^2 A_{\nu}(q_y) \frac{2\nu q_y^2}{\omega^2 + \nu^2 q_y^4} \Biggr\},
    \label{eq: dyfactor infinite}
\end{align}
where $S^{\rm th}_{\mathrm{Ray}, \mathrm{eq}}$ represents the equilibrium contribution from the Rayleigh line, given by
\begin{align}
    S^{\rm th}_{\mathrm{Ray}, \mathrm{eq}} = \frac{c_p-c_v}{c_p} S^{\rm th}_{\mathrm{eq}}.
\end{align}
Here, the nonequilibrium correction is proportional to $(\nabla T)_0^2$, similar to $S(q_y)$.
The coefficients of the nonequilibrium correction, $A_{\nu}(q_y)$ and $A_{T}(q_y)$, exhibits the $q_y^{-4}$ divergence, given by
\begin{align}
    A_{\nu}(q_y) = \frac{a_T}{\nu} A_{T}(q_y) = \frac{k_B c_p}{T_{\mathrm{mid}}(\nu^2 - a_T^2)} \frac{1}{q_y^4}.
    \label{eq: expression for Anu and AT}
\end{align}
Since the static structure factor $S(\bm{q})$ is simply $S(t, \bm{q})$ at $t = 0$, the $q_y^{-4}$ divergence of $A_{\nu}(q_y)$ and $A_{T}(q_y)$ manifests the presence of the LRCs.

To detect this $q_y^{-4}$ divergence in our MD simulations, we fit $S(\omega, q_y)$ to Eq.~(\ref{eq: dyfactor infinite}) and extract $A_{\nu}(q_y)$ and $A_{T}(q_y)$.
The results are shown in Figs.~\ref{fig3}~(c) and (d), where the obtained $A_{\nu}(q_y)$ and $A_{T}(q_y)$ are plotted as a function of $q_y^{-4}$.
The colored dashed lines represent the theoretical prediction Eq.~(\ref{eq: expression for Anu and AT}) with the parameters obtained from the separate MD simulation~\footnotemark[1].
These figures show a good agreement between the MD results and the theoretical prediction, confirming that both $A_{\nu}(q_y)$ and $A_{T}(q_y)$ exhibit a clear $q_y^{-4}$ scaling, revealing the presence of LRCs.

The second protocol aims to demonstrate that the MD results for $S(\omega, \bm{q})$ can be quantitatively described by fluctuating hydrodynamics.
To this end, Figs.~\ref{fig3}~(e) and (f) compare the MD results with the theoretical prediction [Eq.~(\ref{eq: dyfactor infinite}) with Eq.~(\ref{eq: expression for Anu and AT})]. 
We present the results for two values of $q_y$ to provide unambiguous evidence.
These figures show excellent agreement between the MD results and the theoretical prediction.
We stress that no fitting parameters were used in these figures; the parameters in Eqs.~(\ref{eq: dyfactor infinite}) and (\ref{eq: expression for Anu and AT}) were determined from the separate MD simulation~\footnotemark[1].
As shown in Fig.~\ref{fig3}~(b), $S(\omega, \bm{q})$ clearly contains the nonequilibrium contribution. 
The excellent agreement shown in Figs.~\ref{fig3}~(e) and (f) indicates that fluctuating hydrodynamics accurately describes these nonequilibrium effects.
By recalling that fluctuating hydrodynamics predicts the LRCs under temperature gradients [Eq.~(\ref{eq: noneq part of stfactor})], we can interpret this excellent agreement as indirect evidence for the presence of the LRCs.

\begin{figure}[tb]
\begin{center}
\includegraphics[scale=1.0]{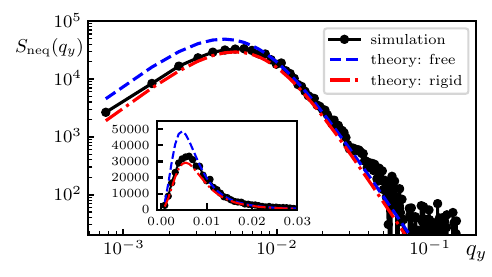}
\end{center}
\vspace{-0.4cm}
\caption{
The finite-size effects on the static structure factor $S(q_y)$, shown as a log-log plot.
The inset shows the same data on a linear scale. 
The parameters are chosen as $(L_x, L_y)=(512, 8192)$ and $(T_{\mathrm{left}}, T_{\mathrm{right}})=(1.5, 2.5)$.
The black curve represents the MD result, which is the same as that in Fig.~\ref{fig2}~(a)
The blue dashed and red dot-dashed curves represent the theoretical predictions with the free~\cite{Ortiz_de_Zarate2001-ho} and rigid boundary conditions~\cite{Ortiz_de_Zarate2001-ab}, respectively. 
Fluctuating hydrodynamics, irrespective of the boundary conditions, accurately reproduces the shape of the $S_{\noneq}(q_y)$ curve obtained from MD simulations.
}
\label{fig4}
\end{figure}
\begin{table*}[tb]
\centering
\begin{tabularx}{1.9\columnwidth}{C{1.5cm}|C{8cm}|C{2.5cm}|C{4cm}} \hline
Approach & Description & Figure & Experimental counterpart \\ \hline
1 & Direct observation of $q_y^{-4}$ divergence in $S(q_y)$ & Fig.~\ref{fig2}~(b) & \cite{Vailati1996-dw, Takacs2011-gg} \\
2 & Observation of $q_y^{-4}$ divergence in $S(\omega, q_y)$ & Figs.~\ref{fig3}~(c) and (d) & \cite{Law1988-zb, Law1990-pg, Segre1992-cd, Li1994-rh, Li1994-hv, Li1994-vw} \\
3 & Comparison of $S(\omega, q_y)$ with fluctuating hydrodynamics & Figs.~\ref{fig3}~(e) and (f) & \cite{Takacs2008-ax, Takacs2011-gg} \\ \hline
\end{tabularx}
\caption{
Summary of the three approaches used to investigate the LRCs.
Each approach has a corresponding experimental counterpart.
A review of early experimental studies can be found in Refs.~\cite{Dorfman1994-cl, De_Zarate2006-xw}.
More recent experimental investigations are reviewed in Refs.~\cite{Vailati2012-yz, Bedeaux2015-lu, Croccolo2016-mn}.}
\label{tab1}
\end{table*}
\sectionprl{Discussions}
We address the finite-size effects observed in Fig.~\ref{fig2}~(b).
These finite-size effects hinder the observation of the LRCs, necessitating large-scale MD simulations.
Indeed, as seen in Fig.~\ref{fig2}~(b), a large-scale simulation with a wall distance $L_x$ of at least $512$ or $1024$ was required to clearly identify the $q_y^{-4}$ scaling of $S(q_y)$.
Similar finite-size effects have been previously reported in theoretical~\cite{Ortiz_de_Zarate2001-ho, Ortiz_de_Zarate2001-ab, Ortiz_de_Zarate2015-pd} and experimental studies~\cite{Mazzoni2006-ji, Croccolo2006-bm, Takacs2011-gg, Vailati2011-rz, Cerbino2015-jh}.

To investigate the origin of these effects, we focus on the MD result for the system with $(L_x, L_y) = (512, 8192)$ and $(T_{\mathrm{left}}, T_{\mathrm{right}})=(1.5, 2.5)$ shown in Fig.~\ref{fig2}~(a).
Figure~\ref{fig4} replots the result of this simulation on a log-log scale (black curve). 
We compare this MD result with the predictions from fluctuating hydrodynamics, explicitly incorporating the finite-size effects through the application of boundary conditions.
We consider two types of boundary conditions: a free boundary condition~\cite{Ortiz_de_Zarate2001-ho} and a rigid boundary condition~\cite{Ortiz_de_Zarate2001-ab}.
Ortiz de Z\'{a}rate et al. analyzed fluctuating hydrodynamics incorporating these boundary conditions and provided theoretical expressions for $S(\bm{q})$ in Refs.~\cite{Ortiz_de_Zarate2001-ho, Ortiz_de_Zarate2001-ab}.

Figure~\ref{fig4} reveals two principal mechanisms responsible for the observed finite-size effects.
First, the elongated geometry of our system ($L_x \ll L_y$) restricts the development of the LRCs.
This is evident from the remarkable agreement between the shape of the $S_{\noneq}(q_y)$ curve obtained from MD simulations and those predicted by fluctuating hydrodynamics, irrespective of the boundary conditions.
The theoretical prediction indicates that the maximum intensity occurs at approximately $q_{y} \simeq 1/L_x$~\cite{Ortiz_de_Zarate2001-ho, Ortiz_de_Zarate2001-ab}.
Therefore, this correspondence of the peak positions suggests that a sufficiently large wall distance $L_x$ is necessary to induce observable LRCs.
Second, the specific boundary conditions applied to the velocity field further suppress the emergence of the LRCs.
We find that the MD result is roughly $30\%$ lower than the prediction for the free boundary condition.
This discrepancy highlights the suppression of the LRCs by the boundary conditions and suggests that designing boundary conditions could be a strategy for observing stronger LRCs.

\sectionprl{Concluding remarks}
In this study, we performed MD simulations to investigate the presence of the LRCs in fluids under temperature gradients.
Our simulations employed the three distinct approaches, each with a corresponding experimental counterpart, as summarized in Table~\ref{tab1}.
First, we directly observed the characteristic $q_y^{-4}$ divergence in the static structure factor, $S(q_y)$, consistent with the experimental observations by Vailati et al.~\cite{Vailati1996-dw} and Takacs et al.~\cite{Takacs2011-gg}.
This approach provides direct confirmation of the existence of the LRCs.
Second, we examined the dynamic structure factor $S(\omega, q_y)$ and detected the same $q_y^{-4}$ divergence as in $S(q_y)$.
The corresponding experimental observation has been extensively performed since the 1980s~\cite{Law1988-zb, Law1990-pg, Segre1992-cd, Li1994-rh, Li1994-hv, Li1994-vw}.
Finally, we compared our MD results for $S(\omega, q_y)$ with the predictions from fluctuating hydrodynamics and found excellent quantitative agreement.
This agreement further validates the existence of the LRCs, whose approach was used experimentally by Takacs et al.~\cite{Takacs2008-ax, Takacs2011-gg}.
In addition, we observed the similar finite-size effects as those observed in previous experiments~\cite{ Mazzoni2006-ji, Croccolo2006-bm, Takacs2011-gg, Vailati2011-rz, Cerbino2015-jh}
From these observations, we conclude that MD simulations are a powerful complementary tool to experiments for investigating LRCs in realistic non-equilibrium situations.

Our simulations focused on a two-dimensional fluid.
While previous experimental observations of LRCs have primarily focused on three-dimensional fluids, the experimental possibility of realizing two-dimensional fluid systems has recently been discussed~\cite{Brogioli2017-hf}, and further developments are anticipated.
It is important to note that transport coefficients in two-dimensional fluids exhibit a logarithmic divergence with system size~\cite{Forster1977-lr, Donev2011-hf, Nakano2025-tj}.
Although we did not observe a significant system-size dependence of these coefficients within the range of parameters explored in our study, this difference could become relevant for more quantitative comparisons between MD simulations and fluctuating hydrodynamics.
It is left for future work.


\begin{acknowledgments}
We thank Hiroshi Watanabe and Kyosuke Adachi for their helpful comments.
The computations in this study were performed using the facilities of the Supercomputer Center at the Institute for Solid State Physics, the University of Tokyo.
The authors are grateful for the fruitful discussions at the workshop “Advances in Fluctuating Hydrodynamics: Bridging the Micro and Macro Scales” hosted by YITP at Kyoto University and RIKEN iTHEMS.
The work of H. N. was supported by JSPS KAKENHI Grants No. JP22K13978.
The work of K. Y. was supported by JST SPRING Grants No. JPMJSP2110.
\end{acknowledgments}


%

\clearpage




\setcounter{equation}{0}
\setcounter{figure}{0}
\setcounter{table}{0}
\setcounter{page}{1}
\renewcommand{\thepage}{S\arabic{page}}  
\renewcommand{\thesection}{S\arabic{section}}   
\renewcommand{\thetable}{S\arabic{table}}   
\renewcommand{\thefigure}{S\arabic{figure}}
\renewcommand{\theequation}{S\arabic{equation}}
\renewcommand{\bibnumfmt}[1]{[S#1]}
\renewcommand{\citenumfont}[1]{S#1}

\begin{widetext}

\begin{center}
{\large \bf Supplemental Material for  \protect \\ 
  ``Long-Range Correlations under Temperature Gradients:\\ A Molecular Dynamics Study of Simple Fluids" }\\
\vspace*{0.3cm}
Hiroyoshi Nakano$^{1}$, and Kazuma Yokota$^{2}$
\\
\vspace*{0.1cm}
$^{1}${\small \it Institute for Solid State Physics, University of Tokyo, Kashiwa, Chiba 277-8581, Japan} \\
$^{2}${\small \it Department of Physics, Kyoto University, Kyoto 606-8502, Japan} 
\end{center}

\section{MD results for a dilute gas}
\label{app1}
We present the MD results for a dilute gas.
The simulation setup is the same as in the main text, but the density is chosen to be $\rho=0.05$.
The fundamental properties of the fluid at this density are summarized in Sec.~\ref{app3}.

\subsection{Static structure factor}
Figure~\ref{supfig1} illustrates how the nonequilibrium correction of the static structure factor $S_{\noneq}(q_y):=(S(q_y)-S_{\eq})/(\nabla T)_0^2$ behaves under different temperature gradients and system sizes.
In Fig.~\ref{supfig1} (a), we keep the system size constant and vary the temperature gradient $(\nabla T)_0$.
As seen in this panel, $S_{\noneq}(q_y)$ exhibits a slight dependence on the temperature gradient. While this dependence is stronger than that observed in the dense liquid [Fig.~2(a)], it is less pronounced than expected, considering that the applied temperature gradients are quite large.
In Fig.~\ref{supfig1} (b), we maintain the temperatures at the two ends constant and vary the wall distance $L_x$.
With increasing $L_x$, the simulation results converge towards the theoretical prediction for an infinitely large system [Eq.~(5)], which is consistent with the case of the dense liquid [Fig.~2(b)].
This convergence confirms that our MD simulations accurately capture the expected $q_y^{-4}$ divergence, providing direct evidence of the LRCs even in the dilute gas.

\subsection{Dynamics structure factor}
Figure~\ref{supfig2} presents a detailed analysis of the dynamical structure factor $S(q_y, \omega)$.
Figure~\ref{supfig2} (a) presents the overall behavior of $S(q_y, \omega)$ for $\omega > 0$.
Compared to the dense liquid, the dilute gas has a larger sound speed, resulting in a smaller separation between the Rayleigh peak ($\omega = 0$) and the Brillouin peaks ($\omega = \pm c_s q_y$). 
Figure~\ref{supfig2} (b) shows an enlarged view of the Rayleigh peak, comparing the nonequilibrium MD result with the corresponding equilibrium result.
We clearly observe the enhancement of the Rayleigh peak due to the temperature gradient.

To quantify this enhancement, we extract $A_T(q_y)$ and $A_{\nu}(q_y)$ by fitting the MD results for the Rayleigh peak with the theoretical prediction for an infinitely large system [Eq.~(13)].
Figures~\ref{supfig2} (c) and (d) show $A_T(q_y)$ and $A_{\nu}(q_y)$ as a function of $q_y^{-4}$.
In the small $q_y$ regime (large $q_y^{-4}$), we observe the good agreement between the MD results and the theoretical prediction [Eq.~(15)], confirming the expected $q_y^{-4}$ divergence of $A_T(q_y)$ and $A_{\nu}(q_y)$.
This is the signature of the LRCs.

However, compared to the dense liquid case, the data in Figure~\ref{supfig2} (c) and (d) exhibit more significant deviations from the theoretical prediction.
This discrepancy can be attributed to two main factors.
First, in the small $q_y$ regime, the Rayleigh and Brillouin peaks are not well-separated, making it challenging to isolate the contribution of the Rayleigh peak.  Second, in the large $q_y$ regime, the enhancement of fluctuations due to the temperature gradient is relatively weak, leading to larger measurement errors.
More refined simulations are required to obtain clearer data.
  
Finally, Figs.~\ref{supfig2} (e) and (f) present a direct comparison between the MD results for $S(q_y, \omega)$ and the theoretical prediction [Eq.~(13)] without any fitting parameters.
The excellent agreement shown in these panels confirms the validity of fluctuating hydrodynamics for describing nonequilibrium fluctuations even in dilute gas systems.

\subsection{Finite-size effects}
Figure~\ref{supfig3} presents the finite-size effects on the static structure factor $S(q_y)$, comparing the MD result with theoretical predictions for free and rigid boundary conditions.
The MD result more closely resembles the prediction for the rigid boundary condition.
This resemblance is stronger than in the dense liquid discussed in the main text.
However, a precise quantitative comparison is difficult due to the limitations of the first-order Galerkin approximation used for the rigid boundary condition.
We can at least conclude that our MD setup does not correspond to a system with free boundaries since the theoretical result for the free boundary condition is exact.

\begin{figure*}[tb]
\begin{center}
\includegraphics[scale=1.0]{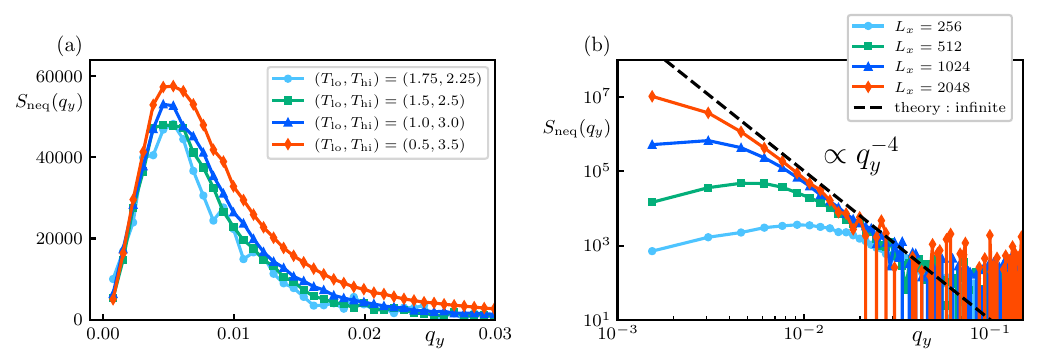}
\end{center}
\vspace{-0.5cm}
\caption{
The MD results of the static structure factor $S(q_y)$, corresponding to Fig.~2 in the main text.
(a) Dependence of $S(q_y)$ on the temperature gradient $(\nabla T)_0$.
We fix the system size at $(L_x, L_y)=(512, 8192)$ and vary the temperatures at the two ends as $(T_{\mathrm{lo}}, T_{\mathrm{hi}})=(1.75, 2.25)$, $(1.5, 2.5)$, $(1.0, 3.0)$, $(0.5, 3.5)$, where the midpoint temperature is fixed at $T_{\mathrm{mid}}:=(T_{\mathrm{lo}}+T_{\mathrm{hi}})/2=2.0$.
(b) Dependence of $S(q_y)$ on the system size $L_x$.
We maintain the temperatures at the two ends at $(T_{\mathrm{lo}}, T_{\mathrm{hi}})=(1.5, 2.5)$ and vary the system size as $(L_x, L_y)=(256, 4096)$, $(512, 4096)$, $(1024, 4096)$, $(2048, 4096)$.
The black dashed line represents the theoretical prediction [Eq.~(5)] for an infinitely large system.
}
\label{supfig1}
\end{figure*}
\begin{figure*}[tb]
\begin{center}
\includegraphics[scale=1.0]{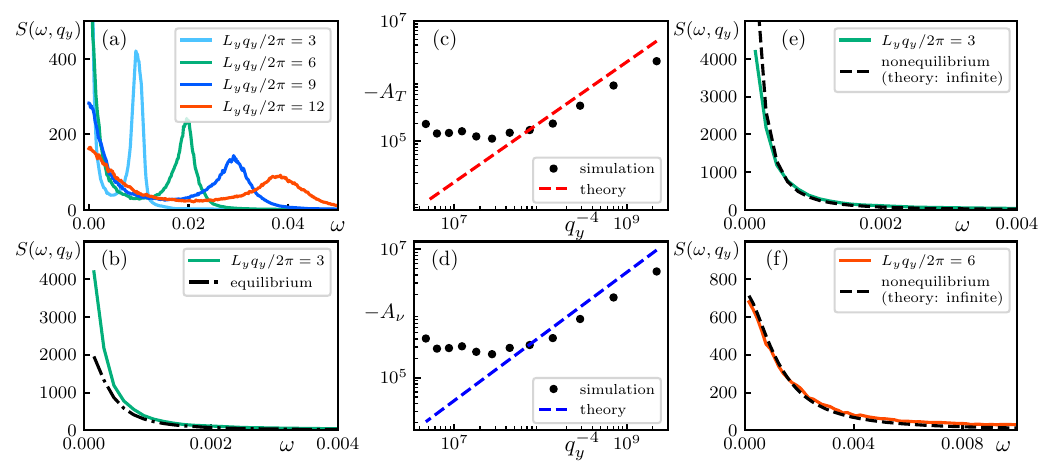}
\end{center}
\vspace{-0.5cm}
\caption{
The MD results of the dynamic structure factor $S(\omega, q_y)$, corresponding to Fig.~3 in the main text.
The parameters are chosen as $(L_x, L_y)=(2048, 4096)$ and $(T_{\mathrm{lo}}, T_{\mathrm{hi}})=(1.5, 2.5)$, corresponding to the red line in Fig.~\ref{supfig2}.
(a) Rayleigh and Brillouin peaks of $S(\omega, q_y)$ for several wavenumbers, $q_y$. 
(b) $S(\omega, q_y)$ near the Rayleigh peak for $q_y/2\pi=8/L_y$.
The colored solid curve represents an enlarged view of the data in panel (a), while the black dot-dashed curve represents the corresponding equilibrium result.
(c, d) The $q_y^{-4}$ dependence of $A_T(q_y)$ and $A_{\nu}(q_y)$.
The colored solid lines represent the theoretical predictions [Eq.~(15)] for an infinitely large system.
(e, f) Comparison of the Rayleigh peak of $S(\omega, q_y)$ with theoretical predictions for $q_y/2\pi = 6/L_y$ and $q_y/2\pi = 12/L_y$.
The colored solid lines represent the MD results; in particular, the data in panel (e) are the same as those in panel (b).
The black dashed curves represent the theoretical predictions [Eq.~(13)] for an infinitely large system. 
}
\label{supfig2}
\end{figure*}
\begin{figure}[tb]
\begin{center}
\includegraphics[scale=1.0]{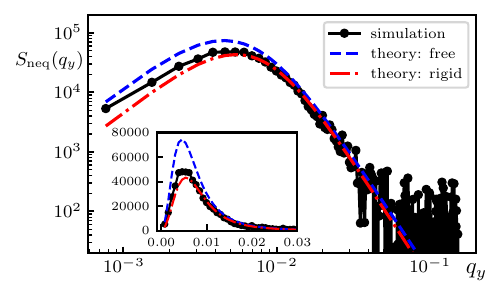}
\end{center}
\vspace{-0.5cm}
\caption{
The finite-size effects on the static structure factor $S(q_y)$, corresponding to Fig.~4 in the main text.
This figure is shown as a log-log plot, while the inset shows the same data on a linear scale. 
The parameters are chosen as $(L_x, L_y)=(512, 8192)$ and $(T_{\mathrm{lo}}, T_{\mathrm{hi}})=(1.5, 2.5)$, corresponding to the green line in Fig.~2(a).
The black curve represents the MD result, which is the same as those in Fig.~1(a).
The blue dashed and red dot-dashed curves represent the theoretical predictions with the free and rigid boundary conditions, respectively. 
}
\label{supfig3}
\end{figure}

\begin{figure}[b]
\begin{center}
\includegraphics[scale=0.9]{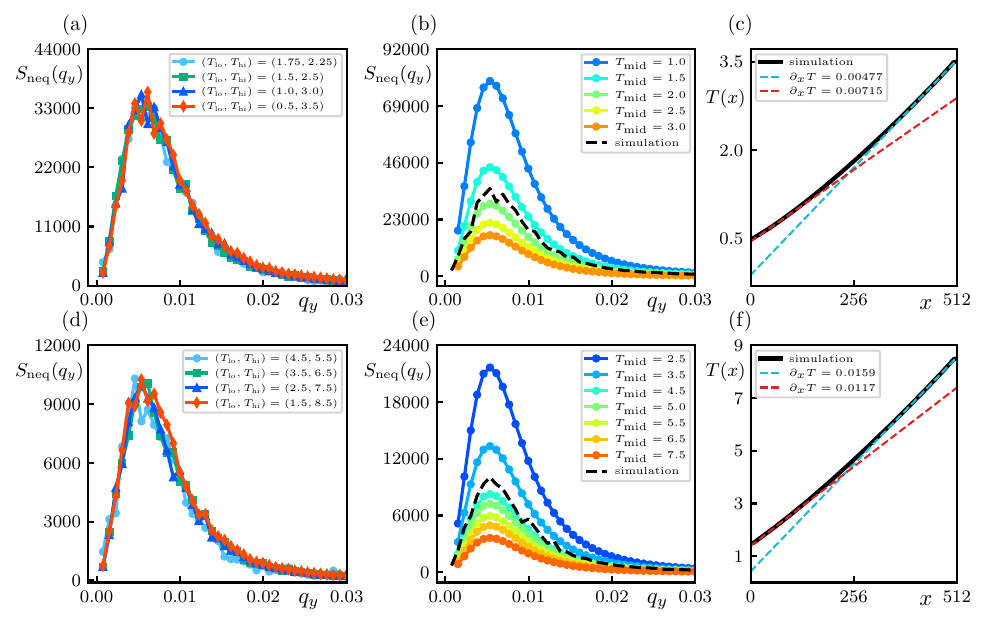}
\end{center}
\vspace{-0.5cm}
\caption{
Temperature Gradient Dependence of $S_{\neq}(q_y)$ observed in MD simulations.
(a)-(c) Results with midpoint temperature fixed at $T_{\rm mid} = 2.0$.
(d)-(f) Results with midpoint temperature fixed at $T_{\rm mid} = 5.0$.
For both sets, the system size is $(L_x, L_y) = (512, 8192)$, and the temperatures at the ends are varied.
(a, d) Dependence of $S_{\neq}(q_y)$ on the temperature gradient $(\nabla T)_0$, obtained from MD simulations.
(b, e) Theoretical predictions for $S_{\neq}(q_y)$, obtained from the linear approximation of fluctuating hydrodynamics with rigid boundary conditions. Thermodynamic and transport properties are evaluated at some temperatures between $T_{\rm mid} = 1.0$ and $T_{\rm mid} = 3.0$ in (b) and between $T_{\rm mid} = 2.5$ and $T_{\rm mid} = 7.5$ in (e).
(c, f) Steady-state temperature profiles, observed in the MD simulations.
}
\label{supfig4}
\end{figure}
\section{Robustness of linear analysis of fluctuating hydrodynamics}

In the main text, we presented the comparison between the results obtained from fluctuating hydrodynamics and MD simulations.
This comparison is performed using the linear solution of fluctuating hydrodynamics.
It is important to note that this linear solution is derived under the assumption of small temperature gradients, where the quantitative behavior of LRCs is characterized by the thermodynamic and transport properties of the fluid evaluated at the midpoint temperature $T_{\rm mid}:=(T_{\rm lo} + T_{\rm hi})/2$. 
Standard fluctuating hydrodynamics analyses inherently rely on this assumption, thus depending solely on $T_{\rm mid}$.

However, as shown in the main text, our MD simulations involve significantly large temperature gradients, leading to spatial variations in the fluid's thermodynamic and transport properties.
Nevertheless, the MD results are effectively characterized by the linear solution of fluctuating hydrodynamics.
This appendix provides a more detailed analysis of this observation.
The specific values of the fluid's thermodynamic and transport properties are detailed in the table provided in the subsequent section.

First, Fig.~\ref{supfig4}(a) replots Fig.~2(a) from the main text, showcasing the static structure factors at the parameters discussed therein.
As previously noted, even with a substantial temperature difference, $S_{\noneq}(q_y)$ demonstrates a collapse onto a universal curve.
This MD simulation imposes a significant temperature gradient, leading to substantial variations in thermodynamic and transport properties across different spatial locations.
To highlight this point, Fig.~\ref{supfig4}(b) presents the theoretical prediction for $S_{\noneq}(q_y)$ obtained from fluctuating hydrodynamics, where the thermodynamic and transport properties are evaluated at specific, representative temperatures within the simulation domain.
These predictions are generated using the linear approximation of fluctuating hydrodynamics under rigid boundary conditions.
We here selected several representative temperatures spanning the range used in our simulations.
For instance, the blue line represents the predicted $S_{\noneq}(q_y)$ using fluid properties evaluated at $T_{\rm mid} = 1.0$, while the orange line corresponds to $T_{\rm mid} = 3.0$.
This figure clearly demonstrates that the theoretical prediction of $S_{\noneq}(q_y)$ exhibits a marked dependence on $T_{\rm mid}$ within the temperature range relevant to our simulations.

Furthermore, to exemplify the spatial variation of thermodynamic and transport properties under the temperature gradient imposed in our simulations, Fig.~\ref{supfig4}(c) presents the steady-state temperature profile.
Notably, the temperature profile exhibits a non-linear shape, which is attributed to the spatial dependence of the thermal conductivity.

From Figs.~\ref{supfig4}(a) and (b), we find the nontriviality of our MD results.
Specifically, a naive expectation would be that the large temperature gradient used in our MD simulations would result in a temperature gradient dependence of $S_{\noneq}(q_y)$.
However, in actuality, no such temperature gradient dependence is observed, and $S_{\noneq}(q_y)$ is robustly described by the linear solution of fluctuating hydrodynamics, even under such significant temperature gradients where local properties vary substantially across the system.

To further validate this finding, we performed additional numerical simulations under even more extreme temperature gradients.
In these simulations, we considered an extreme case: increasing the temperature difference between the edges to $(T_{\rm lo}, T_{\rm hi}) = (1.5, 8.5)$ while maintaining a midpoint temperature of $T_{\rm mid}=5.0$.
The results are depicted in Figs.~\ref{supfig4}(d)-(f), which follow the same arrangement as Figs.~\ref{supfig4}(a)-(c).
Remarkably, even in this extreme case, the LRCs remain well-described by the midpoint temperature, despite the significant variation of local properties across the system.

Thus, our MD simulations demonstrate the surprising robustness of predictions derived from fluctuating hydrodynamics under the small-gradient condition.
The underlying physics remains unclear and will be addressed in future work.

\begin{figure}[b]
\begin{center}
\includegraphics[scale=1.0]{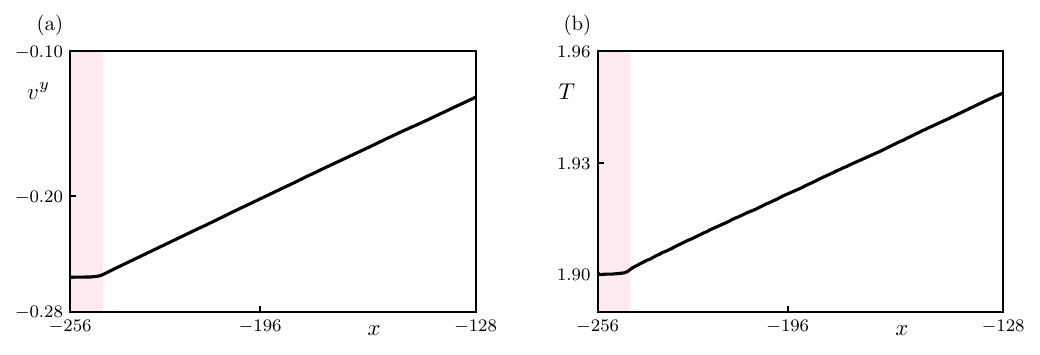}
\end{center}
\vspace{-0.5cm}
\caption{
Typical velocity and temperature profiles near the wall.
The wall is located at $x = -256.0$, and the light red region represents the Langevin thermostat region.
(a) Velocity profile obtained when inducing uniform shear flow.
(b) Temperature profile obtained when applying a temperature gradient.
}
\label{supfig5}
\end{figure}
\begin{table}[b]
\centering
\begin{tabularx}{1\columnwidth}{@{\extracolsep{\fill}}c|c|c|c|c|c}
\hline
Density $\rho$ & System Size $(L_x, L_y)$ & Temperatures $(T_{\rm lo}, T_{\rm hi})$ & Relaxation Loop (steps) & Observation Loop (steps) & Number of Samples \\
\hline
$0.05$ & $(256, 4096)$ & $(1.5, 2.5)$ & $2,000,000$ & $10,000,000$ & 1152 \\
$0.05$ & $(512, 4096)$ & $(1.5, 2.5)$ & $5,000,000$ & $30,000,000$ & 1152 \\
$0.05$ & $(1024, 4096)$ & $(1.5, 2.5)$ & $15,000,000$ & $30,000,000$ & 1440 \\
$0.05$ & $(2048, 4096)$ & $(1.5, 2.5)$ & $20,000,000$ & $40,000,000$ & 2304 \\
$0.05$ & $(512, 8192)$ & $(1.75, 2.25)$ & $5,000,000$ & $40,000,000$ & 288 \\
$0.05$ & $(512, 8192)$ & $(1.5, 2.5)$ & $5,000,000$ & $30,000,000$ & 288 \\
$0.05$ & $(512, 8192)$ & $(1.0, 3.0)$ & $5,000,000$ & $30,000,000$ & 288 \\
$0.05$ & $(512, 8192)$ & $(0.5, 3.5)$ & $5,000,000$ & $30,000,000$ & 288 \\
$0.7$ & $(128, 4096)$ & $(1.5, 2.5)$ & $4,000,000$ & $10,000,000$ & 576 \\
$0.7$ & $(256, 4096)$ & $(1.5, 2.5)$ & $5,000,000$ & $10,000,000$ & 576 \\
$0.7$ & $(512, 4096)$ & $(1.5, 2.5)$ & $10,000,000$ & $10,000,000$ & 288 \\
$0.7$ & $(1024, 4096)$ & $(1.5, 2.5)$ & $15,000,000$ & $20,000,000$ & 288 \\
$0.7$ & $(512, 8192)$ & $(1.75, 2.25)$ & $10,000,000$ & $10,000,000$ & 144 \\
$0.7$ & $(512, 8192)$ & $(1.5, 2.5)$ & $10,000,000$ & $10,000,000$ & 72 \\
$0.7$ & $(512, 8192)$ & $(1.0, 3.0)$ & $10,000,000$ & $10,000,000$ & 72 \\
$0.7$ & $(512, 8192)$ & $(0.5, 3.5)$ & $10,000,000$ & $10,000,000$ & 72 \\
\hline
\end{tabularx}
\caption
{
Relaxation steps, observation steps, and the number of samples used to observe the static structure factor $S(q_y)$ for each set of parameters.
}
\label{suptab1}
\end{table}
\section{Simulation setup and observation protocol}
\label{app2}

In this Appendix, we provide supplementary information regarding the MD simulations. 

\subsection{Simulation setup}

All simulations were carried out using LAMMPS.
We utilized the velocity Verlet algorithm with a time step of $dt=0.002$ for time integration.
The particles are confined by walls at $x = -0.5L_x$ and $x = 0.5L_x$, while the periodic boundary condition is applied along the $y$ axis [Fig.~1(a)].
The walls were defined by the following potentials.
For the wall $x = - 0.5L_x$:
\begin{align}
V_{\mathrm{WALL,-}}(\boldsymbol{r}) =
\begin{cases}
4\epsilon \Big[ \left( \frac{\sigma}{x+0.5L_x} \right)^{12} - \left( \frac{\sigma}{x+0.5L_x} \right)^6 + \frac{1}{4} \Big], & -0.5L_x < x \leq -0.5L_x + 2^{1/6}\sigma , \\
0, & -0.5L_x + 2^{1/6}\sigma < x.
\end{cases}
\end{align}
For the wall $x = 0.5L_x$:
\begin{align}
V_{\mathrm{WALL,+}}(\boldsymbol{r}) =
\begin{cases}
0, & x < 0.5L_x - 2^{1/6} \sigma ,  \\
4\epsilon \Big[ \left( \frac{\sigma}{x-0.5L_x} \right)^{12} - \left( \frac{\sigma}{x-0.5L_x} \right)^6 + \frac{1}{4} \Big], & 0.5L_x - 2^{1/6}\sigma \leq x < 0.5L_x.
\end{cases}
\end{align}
where $\epsilon$ and $\sigma$ represent the well depth and atomic diameter, respectively.

To establish a temperature gradient along the $x$ axis, we applied Langevin thermostats in the regions $[-0.5L_x, -0.48L_x]$ and $[0.48L_x, 0.5L_x]$.
The dynamics of particles in these regions are governed by the following equations:
\begin{align}
    \frac{d \bm{r}_i}{dt} &= \frac{\bm{p}_i}{m} \\
    \frac{d \bm{p}_i}{dt} &= -\sum_{i\neq j}\frac{\partial V_{\rm WCA}(|\bm{r}_i-\bm{r}_j|)}{\partial \bm{r}_i} - \frac{\partial V_{\rm WALL,+}(\bm{r}_i)}{\partial \bm{r}_i} - \frac{\partial V_{\rm WALL,-}(\bm{r}_i)}{\partial \bm{r}_i} - \gamma \bm{p}_i + \bm{\xi}_i(t)
\end{align}
where $\bm{\xi}_i(t)$ represents Gaussian white noise satisfying:
\begin{align}
    \langle \xi^a_i(t) \xi^b_j(t')\rangle = 2k_B T \gamma \delta^{ab} \delta_{ij} \delta(t-t')
\end{align}
The dynamics of particles outside the thermostat regions are described by the purely Hamiltonian equations of motion:
\begin{align}
    \frac{d \bm{r}_i}{dt} &= \frac{\bm{p}_i}{m} \\
    \frac{d \bm{p}_i}{dt} &= -\sum_{i\neq j}\frac{\partial V_{\rm WCA}(|\bm{r}_i-\bm{r}_j|)}{\partial \bm{r}_i} - \frac{\partial V_{\rm WALL,+}(\bm{r}_i)}{\partial \bm{r}_i} - \frac{\partial V_{\rm WALL,-}(\bm{r}_i)}{\partial \bm{r}_i}
\end{align}

In all simulations, the friction coefficient was fixed at $\gamma=1$. Treating the thermostat regions as effective walls, we consider the positions $x= \pm 0.48L_x$ as the boundaries in the hydrodynamic description.
To validate the appropriateness of these wall positions, Fig.~\ref{supfig5} shows typical velocity and temperature profiles obtained when inducing shear flow and a temperature gradient, respectively.
The uniform shear flow is generated by applying constant and opposite external forces $\bm{f}=(0,f)$ and $\bm{f}=(0,-f)$ to the two thermostat regions.
This results in a constant flow velocity of approximately $\pm f/ \gamma$ within the thermostat regions, thereby establishing a uniform shear flow.
The temperature gradient is achieved by setting different temperatures in the two thermostat regions.

The light red region in Fig.~\ref{supfig5} represents the thermostat region $[-0.5L_x, -0.48L_x]$.
Within this region, both the velocity and temperature are maintained at constant values.
This observation justifies treating the thermostat regions as effective walls.
Consequently, the position $x= - 0.48L_x$ is considered as the effective wall position for our hydrodynamic description.



\subsection{Observation protocol}
Our simulation procedure consisted of a relaxation phase and an observation phase.
As an initial condition, we generated a random spatial configuration of particles and assigned velocities drawn from a Maxwell--Boltzmann distribution at temperature $T$.
For larger systems [$(L_x,L_y)=(1024,4096)$ for $\rho=0.7$], as an exception, we first performed a pre-equilibration run, where we relaxed the system to thermal equilibrium by applying the Langevin thermostat with temperature $T$ to all particles.
To relax the system to a nonequilibrium steady state, we performed a relaxation run with sufficiently large steps.
After that, we performed a production run, during which measurements were collected every $250$ steps (equivalently, $250dt = 0.5$ time units).
Moreover, to enhance the statistical accuracy, multiple independent simulations were performed for each set of parameters, each with different initial conditions and random noise realizations
The final results were obtained by averaging over these independent simulations.
Table~\ref{suptab1} provides the relaxation steps, the observation steps, and the number of samples used to observe the static structure factor $S(q_y)$ for each set of parameters.

\begin{table*}[b]
    \centering
    \begin{tabular}{C{1.5cm}C{2cm}C{2cm}C{4cm}C{4.5cm}}
        \toprule
        Density $\rho$ & Temperature $T$ & Pressure $P$ & \begin{tabular}{c}
        Specific heat capacity \\ at constant pressure $c_p$ \end{tabular}& Thermal expansion coefficient $\alpha_p$ \\
        \midrule
        0.05 & 0.5 & 0.0273 & 2.00 & 1.78\\
        0.05 & 1.0 & 0.0544 & 2.00 & 0.870 \\
        0.05 & 1.5 & 0.0813 & 2.01 & 0.520 \\
        0.05 & 2.0 & 0.108 & 2.00 & 0.475 \\
        0.05 & 2.5 & 0.135 & 2.02 & 0.256 \\
        0.05 & 3.0 & 0.162 & 2.03 & 0.188 \\
        0.05 & 3.5 & 0.189 & 2.03 & 0.139 \\\midrule
        0.7 & 0.5 & 2.13 & 2.41 & 0.431 \\
        0.7 & 1.0 & 3.74 & 2.29 & 0.232 \\
        0.7 & 1.5 & 5.17 & 2.26 & 0.161 \\
        0.7 & 2.0 & 6.51 & 2.22 & 0.126 \\
        0.7 & 2.5 & 7.78 & 2.21 & 0.103 \\
        0.7 & 3.0 & 9.00 & 2.15 & 0.089 \\
        0.7 & 3.5 & 10.2 & 2.11 & 0.079 \\\midrule
        0.7 & 4.5 & 12.4 & 2.50 & 0.052 \\
        0.7 & 5.0 & 13.5 & 2.42 & 0.049 \\
        0.7 & 5.5 & 14.6 & 2.49 & 0.043 \\
        0.7 & 6.5 & 16.7 & 2.39 & 0.039 \\
        0.7 & 7.5 & 18.8 & 2.49 & 0.032 \\
        0.7 & 8.5 & 20.8 & 2.55 & 0.028 \\
        \bottomrule\bottomrule
    \end{tabular}
    \begin{tabular}{C{1.5cm}C{2cm}C{2.0cm}C{2.5cm}C{2.8cm}C{2.8cm}||C{2.7cm}}
        \toprule
        Density $\rho$ & Temperature $T$ & Shear viscosity $\eta$ & Kinetic viscosity $\nu := \eta/\rho$ & Thermal conductivity $\kappa$ & Thermal diffusivity $a_{T} := \kappa / \rho c_p$ & Magnitude of LRCs $A_{\rm neq}$\\
        \midrule
        0.05 & 0.5 & 0.197 & 3.94 & 0.835 & 8.35 & 0.0154 \\
        0.05 & 1.0 & 0.280 & 5.60 & 1.19 & 11.9 & 0.00363 \\
        0.05 & 1.5 & 0.347 & 6.94 & 1.49 & 14.8 & 0.00126 \\
        0.05 & 2.0 & 0.410 & 8.20 & 1.71 & 17.2 & 0.00103 \\
        0.05 & 2.5 & 0.462 & 9.24 & 1.98 & 19.6 & 0.000290 \\
        0.05 & 3.0 & 0.499 & 9.98 & 2.16 & 21.3 & 0.000159 \\
        0.05 & 3.5 & 0.546 & 10.92 & 2.30 & 22.7 & 0.0000886 \\\midrule
        0.7 & 0.5 & 1.21 & 1.73 & 5.76 & 3.41 & 0.00530 \\
        0.7 & 1.0 & 1.39 & 1.99 & 7.16 & 4.47 & 0.00186 \\
        0.7 & 1.5 & 1.52 & 2.17 & 8.14 & 5.15 & 0.00103 \\
        0.7 & 2.0 & 1.61 & 2.30 & 8.96 & 5.77 & 0.000682 \\
        0.7 & 2.5 & 1.71 & 2.44 & 9.53 & 6.16 & 0.000501 \\
        0.7 & 3.0 & 1.79 & 2.56 & 10.1 & 6.71 & 0.000382 \\
        0.7 & 3.5 & 1.87 & 2.67 & 10.7 & 7.24 & 0.000304 \\\midrule
        0.7 & 4.5 & 1.99 & 2.84 & 11.6 & 6.63 & 0.000194 \\
        0.7 & 5.0 & 2.07 & 2.96 & 12.0 & 7.08 & 0.000169 \\
        0.7 & 5.5 & 2.12 & 3.03 & 12.5 & 7.17 & 0.000139 \\
        0.7 & 6.5 & 2.22 & 3.17 & 13.1 & 7.83 & 0.000115 \\
        0.7 & 7.5 & 2.31 & 3.30 & 14.0 & 8.03 & 0.0000844 \\
        0.7 & 8.5 & 2.42 & 3.46 & 13.6 & 7.62 & 0.0000789 \\
        \bottomrule
    \end{tabular}
    \caption{Thermodynamic and transport properties of the fluids}
    \label{suptab2}
\end{table*}
\section{Fundamental properties of fluids}
\label{app3}
Table~\ref{suptab2} lists the thermodynamic and transport properties of the fluids.
We summarize below how these properties were measured.




\subsection{Specific heat capacity at constant volume $c_v$}

The specific heat capacity at constant volume $c_v$ is defined by
\begin{align}
    c_v := \frac{1}{N}\Biggl(\frac{\partial E}{\partial T}\Biggr)_{V}.
    \label{eq:def of cv}
\end{align}
where $E$ is the energy, and $N$ is the particle number.
To determine $c_v$ in our atomic systems from Eq.~(\ref{eq:def of cv}), we prepare the NVT ensemble with the periodic boundary condition for all directions and observe the energy $E$.
The derivative is calculated by fitting the energy data as a function of temperature with a linear function.

The energy is measured for the system size of $(L_x, L_y) = (512, 512)$ over $64$ or $128$ independent samples.
Each sample undergoes the preliminary run with the "fix nve" command in LAMMPS for the $5000000$ step (i.e. $5000000dt=10000$ time) to reach equilibrium.
After that, we collect the energy data every $100$ step during the $5000000$ step (i.e. $5000000dt=10000$ time).

\subsection{Specific heat capacity at constant pressure $c_p$}
The specific heat capacity at constant pressure $c_p$ is defined by
\begin{align}
    c_p := \frac{1}{N}\Biggl(\frac{\partial H}{\partial T}\Biggr)_{P},
    \label{eq:def of cp}
\end{align}
where $H$ is the enthalpy.
To determine $c_p$ in our atomic systems from Eq.~(\ref{eq:def of cp}), we prepare the NPT ensemble with the periodic boundary condition and observe the enthalpy $H:=E+PV$, where $E$ is the energy and $PV$ is the pressure-volume product.
The derivative is calculated by fitting the enthalpy data as a function of temperature with a linear function.

The enthalpy is measured for the system size of $(L_x, L_y) = (512, 512)$ over $64$ or $128$ independent samples.
Each sample undergoes the preliminary run with the "fix npt" command in LAMMPS for the $4000000$ step (i.e. $4000000dt=8000$ time) to reach equilibrium.
Subsequently, we switch the time evolution to the Hamiltonian dynamics with the "fix nve" command in LAMMPS. 
After an additional $1000000$ step (i.e. $1000000dt=2000$ time), we collect the enthalpy data every $100$ step during the $5000000$ step (i.e. $5000000dt=10000$ time).

\subsection{Thermal expansion coefficient}
Thermal expansion coefficient $\alpha_p$ quantifies how volume changes per unit temperature change, which is mathematically defined by
\begin{align}
    \alpha_p := \frac{1}{V}\Biggl(\frac{\partial V}{\partial T}\Biggr)_P = -\frac{1}{\rho}\Biggl(\frac{\partial \rho}{\partial T}\Biggr)_P
    \label{eq:def of alphap}
\end{align}
where $V$ is the area of the system.
The determination of $\alpha_p$ in our atomic systems is performed simultaneously with the one for specific heat capacity $c_p$.
We measure the area as the function of the temperature and estimate the thermal expansion coefficient $\alpha_p$ from Eq.~(\ref{eq:def of alphap}).

\subsection{Shear viscosity}
The shear viscosity $\eta$ is defined through the constitute equation of fluids
\begin{align}
    \Pi_{ij} = p \delta_{ij} + \rho v_i v_j - \eta \Biggl(\frac{\partial v_j}{\partial x_i} + \frac{\partial v_i}{\partial x_j} - \delta_{ij} \frac{\partial v_k}{\partial x_k} \Biggr) - \zeta \delta_{ij} \frac{\partial v_k}{\partial x_k}
\end{align}
where $\Pi_{ij}$ represents the momentum flux.
To determine the shear viscosity $\eta$ in our atomic systems, we use the simpler form of the constitute equation in the Couette flow setup as
\begin{align}
    \Pi_{xy} = \eta \frac{\partial v^x}{\partial y}
\end{align}

The measurement is performed for the system with $(L_x, L_y) = (512, 512)$.
We perform the $128$ independent runs to take an ensemble average.
Each of the $128$ samples undergoes the relaxation run for the $500000$ step (i.e. $500000dt=10000$ time).
After the system reached the nonequilibrium steady state, we observe the shear stress $\Pi_{xy}(x)$ and the velocity field $v_x(x)$ every $1000$ step during the $500000$ step.
Then, the shear viscosity is then estimated as a ratio of the shear stress to the velocity gradient.

\subsection{Thermal conductivity}
Thermal conductivity $\kappa$ is defined through Fourier's law
\begin{align}
    \bm{q} = -\kappa \frac{\partial T}{\partial \bm{x}},
\end{align}
where $\bm{q}$ represents the heat flux.
To determine $\kappa$ in our atomic systems, we use the same setup as given in the main text (Fig.~1(a)).
Because the heat flows only along the $x$ axis, we can apply the simpler form of Fourier's law
\begin{align}
    q_x = -\kappa \frac{\partial T}{\partial x}.
\end{align}

The measurement is performed for the system with $(L_x, L_y) = (512, 512)$.
We perform the $128$ independent runs to take an ensemble average.
Each of the $128$ samples undergoes the relaxation run for the $10000000$ step (i.e. $10000000dt=10000$ time).
After the system reached the nonequilibrium steady state, we observe the heat flux $q_x(x)$ and the temperature field $T(x)$ every $1000$ step during the $15000000$ step.
Then, the thermal conductivity is then estimated as a ratio of the heat flux to the temperature gradient.

\end{widetext}
\end{document}